\documentclass[onecolumn,secnumarabic,amssymb, amsmath, nofootinbib,tightenlines,
nobibnotes, aps, prl,epsfig]{revtex4}
\usepackage{graphicx}% Include figure files
\usepackage{dcolumn}% Align table columns on decimal point
\usepackage{bm}% bold math
\begin{document}
\preprint{APS/123-QED}
\title{Evolution of entropy at small $x$}% Force line breaks with \\

\author{G.R.Boroun}%
 \email{ boroun@razi.ac.ir }
\affiliation{ Physics Department, Razi University, Kermanshah
67149, Iran}% \textbackslash\textbackslash
\author{Phuoc Ha }
\altaffiliation{pdha@towson.edu}%Lines break automatically or can be forced with \\
\affiliation{Department of Physics, Astronomy and Geosciences,
Towson University, Towson, MD 21252}
%\author{K.Kutak}%
% \email{ kutak@pl }
%\affiliation{ Instytut Fizyki Jadrowej im H. Niewodniczanskiego,
%Radzikowskiego 152, 31-342 Krakow, Poland}%

\date{\today}% It is always \today, today,
\begin{abstract}
%%%%%%%%%%%%%%%%%%%%%%%%%%%%%%%%%%%%%%%%%%%%%%%%%%%%%%%
We explore the evolution of the  Deep Inelastic Scattering (DIS)
entropy, defined as $ S(x,\mu^2) \simeq
\ln[xg(x,\mu^2)]$ at small Bjorken variable $x$, where $\mu$ is the observable scale and
 the gluon distribution $xg(x,\mu^2)$ is
derived from the Dokshitzer-Gribov-Lipatov-Altarelli-Parisi
(DGLAP) evolution equations. We aim to evolve the DIS entropy,
which is not directly observable, using a Laplace transform
technique. This approach allows us to obtain an analytical
solution for the DIS entropy based on known initial gluon
distribution functions. We consider both leading-order (LO) and
higher-order approximations for the DIS entropy, incorporating the
evolved gluon distribution function at the initial scale.
The DIS entropy, influenced by purely gluonic emissions, varies with
higher-order corrections to the running coupling. By comparing theoretical
predictions with charged hadron multiplicity data, we define the evolution.
Additionally, we investigate the derivative of the scaling entropy, modeling it
as a function of the running coupling, to determine the parameter $\lambda$, known as the Pomeron intercept.
We find that the values of $\lambda(x,\mu^2)$
decrease as the order of evolution increases, which is consistent
with the Balitsky-Fadin-Kuraev-Lipatov (BFKL) Pomeron in the LO and
NLO approximations. This investigation provides insights into the dynamics of
Quantum Chromodynamics (QCD) at high energies.\\

%%%%%%%%%%%%%%%%%%%%%%%%%%%%%%%%%%%%%%%%%%%%%%%%%%%%%%%
\end{abstract}
 \pacs{***}%PACS, the Physics and Astronomy
                              %Classification Scheme.
 %Use showkeys class option if keyword
\maketitle
%**********************************************************
%%%%%%%%%%%%%%%%%%%%%%%%%%%%%%%%%%%%%%%%%%%%%%%%%%%%%%%%%%%%%%%%%%%%%%%%%%%%%%%%%%%%%%%%%%%%%
\section{Introduction}
Entropy, which  is an important quantity of a system in
thermodynamics, can be described according to the Boltzmann
entropy\footnote{The Boltzmann entropy describes the disorder or
complexity of the system at the microscopic level.} relation
$S=k_{\mathrm{B}}{\ln}W$, where $W$ denotes the number of
microstates that correspond to the same macroscopic thermodynamic
state, or the Gibbs entropy \footnote{The Gibbs entropy turns into
the Boltzmann entropy if all the microstates have the same
probability.} relation $S=-k_{\mathrm{B}}\sum{p_{i}}{\ln}p_{i}$,
where $p_{i}$ gives the probability to find the system in the
state $|i>$ \cite{Kutak1, Wang}. In Deep inelastic Scattering
(DIS), at high-energy interaction, the interaction time between
the virtual photon and the proton is in general much shorter than
the characteristic time scale $t_{n}{\sim}1/E_{n}$. Indeed, the
probe to read off the information about the phases $\phi_{n}$ of
the individual Fock states $|n>$ with $n$ partons in high-energy
interaction is impossible and it causes the resulting information
scrambling \cite{Kutak2}. The mixed state produced from the proton
after the DIS measurement is due to the uncertainty relation
between the phase and the occupation number of the Fock states
where determine the entropy of the multi-hadron state created in
DIS. The entropy of the partons in a DIS experiment resolved by
the authors in Refs.\cite{Kharzeev, Levin} by the following form
\begin{eqnarray}\label{Entropy1_eq}
S_{\mathrm{DIS}}={\ln}N(x,\mu^2),
\end{eqnarray}
where $N(x,\mu^2)$ is the number of partons in a hadron with
longitudinal light-front momentum fraction $x$ of the struck
parton in the target hadron and the observable scale $\mu$ can be
identified with the proton virtuality $\mu^2=Q^2$ in the DIS
measurement where $q^2=-Q^2$ is the momentum transfer
\cite{Brodsky}.  $N(x,\mu^2)$, which represents the number of degrees of freedom in the DIS
measurement, is defined as the total number of partons per
$\ln{1/x}$ for the universal entanglement entropy. The
entanglement entropy is suggested to be
\begin{eqnarray}\label{Numan1_eq}
S(x,\mu^2)={\ln}N(x,\mu^2),
\end{eqnarray}
where
\begin{eqnarray} \label{Numan2_eq}
N(x,\mu^2){\equiv}x\Sigma(x,\mu^2)+xg(x,\mu^2),
\end{eqnarray}
 with \label{Singlet1_eq}
\begin{eqnarray}
x\Sigma(x,\mu^2)=\sum_{f}\bigg{(}q_{f}(x,\mu^2)+\overline{q}_{f}(x,\mu^2)\bigg{)},
\end{eqnarray}
where $g(x,\mu^2)$ and $q_{f}(x,\mu^2)$ denote the parton
densities of the gluon and the quark of flavor $f$, respectively.
As an alternative, the partonic entropy model has been extended in
Refs.\cite{Kharzeev, Levin, Kutak3} based on the dipole entropy
and the von Neumann entropy at small $x$, respectively.\\
In Ref.\cite{Kharzeev}, it is revealed that in the small $x$
region,
 the DIS entropy  is related to the gluon distribution
 $xg(x,\mu^2)$. The relationship between the von Neumann
 entropy\footnote{In analogy to classical (Boltzmann) entropy,
 the quantum (von-Neumann) entropy of a state described by the density operator $\rho$ is
given by the expectation value of the trace of the statistical
operator $S=-\mathrm{tr}[\rho{\ln}\rho]$. The density matrix
$\rho$ is defined as $\rho=|\psi><\psi|$, where $|\psi>$
represents a pure quantum mechanical state. A pure state $|\psi>$,
similar to an elementary particle, has a
quantum entropy of 0.\\
In the proton$^,$s rest frame (where it is in a pure quantum
mechanical state), the DIS probes the spatial region A, which is
only a part of the proton$^,$s wave function. The inclusive DIS
measurement sums over the unobserved part of the wave function
localized in the region B, which is complementary to A. The
reduced density matrix $\rho_{A}=\mathrm{tr}_{B}\rho$ is accessed
rather than the entire density matrix $\rho$. For very slow
partons, gluons are dominant, leading to $S=\ln{N(x)}$, where
$N(x)$ is the number of gluons with longitudinal momentum fraction
$x$ \cite{Kharzeev, Brodsky}.} and the gluon distribution
 accessed at small $x$
in DIS can be expressed in the following form
\begin{eqnarray} \label{Entropyg_eq}
S(x,\mu^2){\simeq}{\ln}\bigg{[}xg(x,\mu^2)\bigg{]},
\end{eqnarray}
where the $\mu$- dependent gluon distribution is derived from the
Dokshitzer-Gribov-Lipatov-Altarelli- Parisi (DGLAP) evolution
equations \cite{DGLAP1, DGLAP2, DGLAP3}. In Quantum Chromodynamics
(QCD), the gluon distribution at a specfic virtuality scale
$\mu^2$ is determined through DGLAP evolution  based on the
behavior of the DIS cross sections $\gamma^{*}+p{\rightarrow}X$.
We wish to evolve the DIS entropy, which is not a directly
observable quantity, using a Laplace transform technique. This
allows us to obtain an analytical method for the solution of the
DIS entropy in terms of known initial gluon distribution
functions. We consider both leading-order (LO) and higher-order
approximations for the DIS entropy, incorporating the evolved
gluon distribution function at the initial scale. The DIS entropy
 is based on the treatment of purely gluonic emissions, which naturally
increase and decrease with higher-order corrections in the running coupling.
By comparing predictions with data for charged hadron multiplicities,
one can clearly define the evolution.\\
In the following, the scaling entropy determines the parameter
$\lambda$, which is the hard Pomeron intercept. In saturation physics, the parameter
$\lambda$  predicts the transverse
momentum-dependent gluon distribution, which grows rapidly as
${\sim}x^{-\lambda}$. In the DIS entropy, various
corrections are employed to extract $\lambda(x,Q^2)$ at
higher-order corrections, considering its dependence
on $x$ at small $x$. We investigate the derivative of the
scaling entropy as a model dependent on the running coupling to
determine the parameter $\lambda(x,Q^2)$, offering a perspective
on the dynamics of QCD at high energies.

The paper is organized as follows: in Sec. II, using the Laplace transform
technique, we show the evolution of both leading-order (LO) and
higher-order approximations for the DIS entropy, incorporating the
evolved gluon distribution function at the initial scale.
Our results of the gluon entropy, evaluated in this work based
on the parametrization
groups, and the investigation of the derivative
of the scaling entropy as a model dependent on the running
coupling to determine the parameter $ \lambda$, the Pomeron
intercept, are presented in Sec.  III.
Finally, the conclusions are given in Sec. IV.

\section{Evolution}

The transverse and longitudinal structure functions are expressed
solely through the singlet quark and gluon densities. This is
because at low values of $x$, the non-singlet quark distributions
become negligibly small compared to the singlet distributions.
This expression is
\begin{eqnarray} \label{Singlet_eq}
F_{k=2,L}(x,\mu^2)=<e^2>\sum_{a=s,g}[B_{k,a}(x){\otimes}xf_{a}(x,\mu^2)],
\end{eqnarray}
where $<e^2>$ represents the average charge squared,
$B_{k,a}(x)$ are the known Wilson coefficient functions, and $\mu$
is the observable scale. The
symbol $\otimes$ denotes the convolution formula.
According to the
DGLAP $\mu^2$-evolution equations, the leading twist gluon
distribution, which dominates at small $x$, evolves according to the following expression \cite{Kotikov}
\begin{eqnarray}\label{DGLAP1_eq}
\frac{{\partial}[xg(x,\mu^{2})]}{{\partial}{\ln}\mu^{2}}{\simeq}-\frac{\alpha_{s}(\mu^2)}{8\pi}
\bigg{(}P_{gg}^{LO}(x)+\frac{\alpha_{s}(\mu^2)}{4\pi}\widetilde{P}_{gg}^{NLO}(x)+...\bigg{)}{\otimes}xg(x,\mu^2).
\end{eqnarray}
Within pQCD, the splitting functions are defined in terms of the
coefficient functions $B_{k,a}(x)$, as
$\widetilde{P}_{gg}^{NLO}(x)={P}_{gg}^{NLO}(x)-B_{2,g}^{NLO}(x){\otimes}{P}_{gg}^{NLO}(x)$.
 In the color dipole model
\cite{Nikolaev1}, the photon wave function depends on the mass of
the quarks in the $q\overline{q}$ dipole, so the contribution to
the structure function $F_{2}$ is defined into the individual
quark flavour pairs, $F_{2}=F^{l}_{2}+F^{c}_{2}$ where $F^{l}_{2}$
refers to the light quark pairs and $F^{c}_{2}$ refers to the
contribution $c\overline{c}$ pair \cite{Golec}. In such a case,
contributions depend on the mass of the quarks by modifying the
Bjorken variable $x$ in the DIS entropy
 $x{\rightarrow}\widetilde{x}_{f}{\equiv}x(1+4m_{c}^{2}/\mu^2)$
  with $m_{c}=1.29^{+0.077}_{-0.053}~\mathrm{GeV}$ where the uncertainties
are obtained through adding the experimental fit, model and
parametrization uncertainties in quadrature \cite{HZ, H1}.\\

In this section, we explore the evolution of the  DIS
entropy, defined as $ S(x,\mu^2) \simeq \ln[xg(x,\mu^2)] =
{\ln}{[}G(x,\mu^2){]}$ at small values of the Bjorken variable $x$, where $G(x,\mu^2)$ denotes
 the gluon distribution. Neglecting the the quark contribution, the
DGLAP evolution equation for the gluon distribution function $G(x,\mu^{2})$ at small $x$ simplifies to:
\begin{eqnarray} \label{Entropy2_eq}
\frac{{\partial}
G(x,\mu^{2})}{{\partial}{\ln}\mu^{2}}{\simeq}{\int_{x}^{1}}
P_{gg}(z, a_{s}(\mu^2)) G(\frac{x}{z},\mu^{2})dz,
\end{eqnarray}
where  $P_{gg}(z, a_{s}(\mu^2))$ is the splitting function defined
by  Ref.\cite{Vogt}
\begin{eqnarray}\label{Coupling_eq}
 P_{gg}(z,a_{s}(\mu^2))&=&\sum_{n=0}
\left(\frac{\alpha_{s}(\mu^2)}{4\pi}\right)^{n+1}
P^{(n)}_{gg}(z)=\sum_{n=0} \left(a_{s}(\mu^2)\right)^{n+1}
P^{(n)}_{gg}(z),
\end{eqnarray}
with $a_{s}(\mu^2) = \frac{\alpha_{s}(\mu^2)}{4\pi}$. The running coupling in the renormalization group equation (RGE)
reads
\begin{eqnarray} \label{Alphas_eq}
\mu^{2}\frac{d\alpha_{s}(\mu^2)}{d\mu^2}=-\bigg{(}b_{0}\alpha^{2}_{s}(\mu^2)+b_{1}\alpha^{3}_{s}(\mu^2)
+b_{2}\alpha^{4}_{s}(\mu^2)+... \bigg{)}
\end{eqnarray}
where $b_{0}=\frac{33-2n_{f}}{12\pi}$ is referred to as the 1-loop
$\beta$-function coefficient, the 2-loop coefficient is
$b_{1}=\frac{153-19n_{f}}{24\pi^{2}}$, and the 3-loop coefficient
is
$b_{2}=\frac{2857-\frac{5033}{9}n_{f}+\frac{325}{27}n_{f}^{2}}{128\pi^{3}}$
for the SU(3) color group.\\

We use the method developed in detail in Refs.\cite{Block1,
Block2, Block3, Block4, BH1, BH2, Boroun1} to obtain the evolution of the gluon distribution
via a Laplace-transform approach, and subsequently determine the evolution of the DIS entropy.
Introducing the variable changes $\upsilon={\ln}(1/x)$ , $w={\ln}(1/z)$, and using
the notations
$\widehat{G}(\upsilon,\mu^2){\equiv}G(e^{-\upsilon},\mu^2)$  and
$\widehat{H}_{gg}(\upsilon,a_{s}(\mu^2))=e^{-\upsilon}
P_{gg}(e^{-\upsilon}, a_{s}(\mu^2))=\sum_{n=0}a^{n+1}_{s}(\mu^2)
\widehat{ P}^{(n)}_{gg}(\upsilon)$, we can rewrite the DGLAP
evolution for the gluon distribution in terms of the convolution
integral
\begin{eqnarray} \label{DGLAP2_eq}
\frac{\partial}{\partial{\ln{\mu^2}}}{\widehat{G}}(\upsilon,\mu^2)&{=}&
\int_{0}^{\upsilon}\widehat{H}_{gg}(\upsilon-w,a_{s}(\mu^2))\widehat{G}
\left(w,\mu^2\right)dw . \end{eqnarray}
By Laplace transforming  Eq. (\ref{DGLAP2_eq}) from  $\upsilon$
to $s$ space and using the fact that
\begin{eqnarray} \label{Laplace1_eq}
\mathcal{L}\bigg{[}\int_{0}^{\upsilon}
\widehat{H}_{gg}(\upsilon-w,\alpha_{s}(\mu^2)) {\widehat{G}(w,
\mu^2)} dw;s\bigg{]}=\mathcal{L}
[\widehat{H}_{gg}(\upsilon,\alpha_{s}(\mu^2));s]{\times}
\mathcal{L}[{\widehat{G}(\upsilon, \mu^2)};s],
\end{eqnarray}
we can write the DGLAP evolution for the gluon distribution in $s$- space as
\begin{eqnarray} \label{Laplace1_eq}
\frac{\partial}{\partial{\ln{\mu^2}}}{g}(s,\mu^2)&{=}&
h_{gg}(s,a_{s}(\mu^2)) \, g \left(s,\mu^2\right) ,
 \end{eqnarray}
where
\footnote{Defining the Laplace transform $S (s,\mu^2) =  \mathcal{L} [
{\widehat{S}(\upsilon, \mu^2)};s] =\mathcal{L} [ S(e^{-\upsilon},\mu^2);s]$,
we have
$$
 S (s,\mu^2) =  \mathcal{L} [ {\widehat{S}(w, \mu^2)};s] =
\int_{0}^{\infty} dw e^{-s {\upsilon}} {\widehat{S}(w, \mu^2)}
$$
and
$$
{\ln}[ g (s,\mu^2)] =  {\ln}[ \mathcal{L} [ {\widehat{G}(w,
\mu^2)};s]] = {\ln} \int_{0}^{\infty} dw e^{-s {\upsilon}}
e^{\widehat{S}(w, \mu^2)} \neq S (s,\mu^2) .
$$
Therefore, $g (s,\mu^2) \neq e^{S (s,\mu^2)}$ and
the DGLAP evolution for the DIS entropy cannot be obtained in $s$- space at this stage.
}
\begin{eqnarray} \label{Laplace2_eq}
g (s,\mu^2)& = & \mathcal{L} [ {\widehat{G}({\upsilon}, \mu^2)};s] , \\
h_{gg}(s,a_{s}(\mu^2)) &=& \mathcal{L}
[\widehat{H}_{gg}(\upsilon,\alpha_{s}(\mu^2));s] = \sum_{n=0} h^{(n)}_{gg}(s,a_{s}(\mu^2)) =\sum_{n=0} a^{n+1}_{s}(\mu^2) h^{(n)}_{gg}(s) , \\
 h^{(n)}_{gg}(s)& =&  \mathcal{L} [\widehat{P}^{(n)}_{gg} (\upsilon);s]= \int_{0}^{\infty}
\widehat{P}^{(n)}_{gg}(\upsilon)e^{-s\upsilon}d\upsilon .
\end{eqnarray}
The solution of  Eq.~ (\ref{Laplace1_eq}) in $s$- space is
\begin{eqnarray} \label{Solution1_eq}
g(s,\mu^2) &=& e^{\sum_{n=0}h^{(n)}_{gg}(s) \tau^{n+1} }
g(s,\mu_0^2) ,
\end{eqnarray}
where
\begin{eqnarray} \label{Tau_eq}
 \tau^{n+1} (\mu^2,\mu_0^2) &=& \int_{\mu_0^2}^{\mu^2}
(\frac{\alpha_{s}(t)}{4\pi})^{n+1} \, d \ln t  .
 \end{eqnarray}
Defining
\begin{eqnarray} \label{Coef2 eq}
K_{GG}(\upsilon, \tau) =  \mathcal{L} ^{-1}
[e^{\sum_{n=0}h^{(n)}_{gg}(s) \tau^{n+1} } ; \upsilon]
=\mathcal{L} ^{-1} [ k_{gg}(s,\tau) ; \upsilon] ,
\end{eqnarray}
we can now express the solution for $\widehat{G}$  in $\upsilon$- space in
terms of the convolution integral
\begin{eqnarray} \label{Gluonevlo1 eq}
 \widehat{G}(\upsilon, \mu^2) &=& \int_0^{\upsilon}
K_{GG}(\upsilon - w, \tau (\mu^2,\mu_0^2) )
\widehat{G}(w,\mu_0^2 ) dw .
  \end{eqnarray}
Alternatively, the solution for $\widehat{G}$ solution in $\upsilon$-space can also be written as
\begin{eqnarray} \label{Gluonevlo3 eq}
\widehat{G}(\upsilon, \mu^2)  = \mathcal{L} ^{-1}[ \{ k_{gg}(s,\tau) g(s,\mu_0^2) \} ; \upsilon] .
 \end{eqnarray}
Transforming the equation (\ref{Gluonevlo1 eq}) back to $x$-space, we have
\begin{eqnarray} \label{Gluonevlo2 eq}
G(x, \mu^2) &=& \int_x^{1} K^{(n)}_{GG}(x - z, \tau
(\mu^2,\mu_0^2) ) G(z,\mu_0^2 ) \frac{dz}{z} . \end{eqnarray}
Since $G(x,\mu^2) \simeq e^{S(x,\mu^2)}$, we obtain the evolution of the DIS entropy as
\begin{eqnarray} \label{Entropy1 eq}
 S(x, \mu^2) &=& \ln \bigg{[} \int_x^{1} K_{GG}(x - z, \tau
(\mu^2,\mu_0^2) ) e^{S(z,\mu_0^2 )} \frac{dz}{z} \bigg{]} .
\end{eqnarray}

To use our solution in the integral representation of Eq. (\ref{Entropy1 eq}), one can first
invert the Laplace transform of the function $k_{gg}$, whose behavior resembles that
 of Dirac $\delta$ function for small $\tau$ -- a formidable task.
In Ref. \cite{Block5}, the authors developed a numerical method for inverse Laplace transforms,
which they used to obtain gluon distributions from the proton structure
function. They also presented a numerical solution for the inverse
Laplace transform of the LO function  $k_{gg}$. This method is general and can be applied to higher-order approximation cases.

Retaining only the ${1}/{s}$ terms in the high-energy region of the
coefficients $h^{(n)}_{gg}(s)$,  we find that \footnote{$\mathrm{BesselI}(1,x)$ represents
the Bessel function $I_{1}(x)$.}
\begin{eqnarray} \label{Bessel eq}
K_{GG}(\upsilon, \tau) =
\delta(\upsilon)+\frac{\sqrt{\xi}}{\sqrt{\upsilon}}\mathrm{BesselI}(1,2\sqrt{\xi}\sqrt{\upsilon})
,
\end{eqnarray}
where $\xi=\sum_{n=0}c^{(n)}\tau^{n+1}$ and the
$c^{(n)}$ are the coefficients of the $1/s$ terms of the splitting functions  in the
$s$-space. We thus find that the evolution of the DIS entropy is given by
\begin{eqnarray} \label{Entropy2 eq}
{S}(x,
\mu^2)&=&\ln{\bigg{[}}G(x,\mu_{0}^2)+\int_{x}^{1}G(z,\mu_{0}^2)
\frac{\sqrt{\xi}}{\sqrt{{\ln}\frac{z}{x}}}\mathrm{BesselI}(1,2\sqrt{\xi}\sqrt{{\ln}\frac{z}{x}})\frac{dz}{z}\bigg{]}\nonumber\\
&&=\ln{\bigg{[}}e^{S(x,\mu_0^2
)}+\int_{x}^{1}e^{S(z,\mu_0^2 )}
\frac{\sqrt{\xi}}{\sqrt{{\ln}\frac{z}{x}}}\mathrm{BesselI}(1,2\sqrt{\xi}\sqrt{{\ln}\frac{z}{x}})\frac{dz}{z}\bigg{]},
\end{eqnarray}
where
\begin{eqnarray}\label{EntropyQ0 eq}
S{(x,\mu_{0}^2)}{\simeq}{\ln}\bigg{[}G(x,\mu_{0}^2)\bigg{]}={\ln}\bigg{[}xg(x,\mu_{0}^2)\bigg{]}.
\end{eqnarray}

By retaining only the ${1}/{s^2}$ terms in the high-energy region of the
coefficients $h^{(n)}_{gg}(s)$,  we find that
\begin{eqnarray} \label{s2term}
K_{GG}(\upsilon, \tau) =
\delta(\upsilon)+\frac{d}{d{\upsilon}}\mathrm{H}(\eta{\upsilon}^2)
,
\end{eqnarray}
where the function $ \mathrm{H}(z)$ is given by
\begin{eqnarray} \label{H-func}
\mathrm{H}(z) = \sum_{n=1}^\infty \frac{z^n}{n!(2n)!}
\end{eqnarray}
and $\eta=\sum_{n=1}d^{(n)}\tau^{n+1}$. Here, the
$d^{(n)}$ are the coefficients of the $1/s^2$ terms of the splitting functions  in the
$s$-space. It follows that the DIS entropy evolves according to:
\begin{eqnarray} \label{Entropy3 eq}
{S}(x,
\mu^2)&=&\ln{\bigg{[}}G(x,\mu_{0}^2)+\int_{x}^{1}G(z,\mu_{0}^2)
\frac{d}{d({\ln}\frac{z}{x})}\mathrm{H}(\eta \, {({\ln}\frac{z}{x}})^2)\frac{dz}{z}\bigg{]}\nonumber\\
&&=\ln{\bigg{[}}e^{S(x,\mu_0^2
)}+\int_{x}^{1}e^{S(z,\mu_0^2 )}
\frac{d}{d({\ln}\frac{z}{x})}\mathrm{H}(\eta {({\ln}\frac{z}{x}})^2)\frac{dz}{z}\bigg{]}
,
\end{eqnarray}
where $S{(x,\mu_{0}^2)}$ is defined by Eq. (\ref{EntropyQ0 eq}).\\

$\bullet$ $\mathbf{LO~ analysis}$:\\

The LO coefficient $h^{(0)}_{gg}(s)$ is given by
\begin{eqnarray} \label{Coef1_eq}
h^{(0)}_{gg}(s)&=&\frac{33-2n_{f}}{3}+12\left[\frac{1}{s}-\frac{2}{s+1}
+\frac{1}{s+2}-\frac{1}{s+3}-\psi(s+1)-\gamma_{E} \right].
 \end{eqnarray}
Here  $\psi(s)$ is the digamma function and $\gamma_{E}= 0.5772156
. . .$ is Euler's constant.  In the limit $s{\rightarrow}0$, $h^{(0)}(s)|_{s{\rightarrow}0}{\simeq}\frac{12}{s}$. \\

At the LO approximation, we rewrite $\xi$ as
\begin{eqnarray}\label{XiLO eq}
\xi^{\mathrm{LO}}{\simeq}\frac{3}{\pi}
\int_{\mu_0^2}^{\mu^2}{\alpha_{s}(t)} \, d \ln t.
\end{eqnarray}
Therefore, the DIS entropy, depending on the initial conditions,
at the LO approximation is given by
\begin{eqnarray} \label{EntropyLO eq}
{S}^{\mathrm{LO}}(x,
\mu^2)&=&\ln{\bigg{[}}G^{\mathrm{LO}}(x,\mu_{0}^2)+\int_{x}^{1}G^{\mathrm{LO}}(z,\mu_{0}^2)
\frac{\sqrt{\xi^{\mathrm{LO}}}}{\sqrt{{\ln}\frac{z}{x}}}\mathrm{BesselI}(1,2\sqrt{\xi^{\mathrm{LO}}}\sqrt{{\ln}\frac{z}{x}})\frac{dz}{z}\bigg{]}.
\end{eqnarray}
The DIS entropy at the initial gluon distribution in the scale
$\mu_{0}^{2}$ is defined by the following form
\begin{eqnarray}\label{EntropyLOQ0 eq}
S^{\mathrm{LO}}(x,\mu_{0}^2){\simeq}{\ln}\bigg{[}xg^{\mathrm{LO}}(x,\mu_{0}^2)\bigg{]}.
\end{eqnarray}

$\bullet$ $\mathbf{NLO~ analysis}$:\\

The evolution of the DIS entropy at the next-to-leading-order
approximation (NLO) is captured in the coefficient of
$h^{(1)}(s)$ which is fully derived in $s$-space by
authors in Ref. \cite{khanpour}.   In the limit $s{\rightarrow}0$,  the largest
terms are
\begin{eqnarray}
h^{(1)}(s)|_{s{\rightarrow}0}{\simeq}\bigg{(}\frac{4}{3}C_{F}T_{f}-\frac{46}{9}C_{A}T_{f}\bigg{)}\frac{1}{s}-(C^{2}_{A}{\ln}(16))\frac{1}{s^{2}},
\end{eqnarray}
with $C_{F}=\frac{N_{c}^{2}-1}{2N_{c}}$, $C_{A}=N_{c}$,
$T_{R}=\frac{1}{2}$, and $T_{f}=T_{R}n_{f}$ for the SU(3) gauge
group, where $C_{A}$ and $C_{F}$ are the color Cassimir operators.
Here the $n_{f}$ is the number of active quark flavors.

Retaining the terms $1/s$  in the limit $s{\rightarrow}0$ and
rewriting $\xi$ as
\begin{eqnarray}\label{XiNLO eq}
\xi^{\mathrm{NLO}}{\simeq}
\int_{\mu_0^2}^{\mu^2}\bigg{[}\frac{3}{\pi}{\alpha_{s}(t)}-
\frac{61}{36\pi^2}{\alpha^{2}_{s}(t)}\bigg{]}\, d \ln t,
\end{eqnarray}
We find that the DIS entropy, depending on the initial conditions, is given at the NLO approximation by
\begin{eqnarray} \label{EntropyNLO eq}
{S}^{\mathrm{NLO}}(x,
\mu^2)&=&\ln{\bigg{[}}G^{\mathrm{NLO}}(x,\mu_{0}^2)+\int_{x}^{1}G^{\mathrm{NLO}}(z,\mu_{0}^2)
\frac{\sqrt{\xi^{\mathrm{NLO}}}}{\sqrt{{\ln}\frac{z}{x}}}\mathrm{BesselI}(1,2\sqrt{\xi^{\mathrm{NLO}}}\sqrt{{\ln}\frac{z}{x}})\frac{dz}{z}\bigg{]}.
\end{eqnarray}

By retaining only the terms $1/s^2$  in the limit $s{\rightarrow}0$ and
rewriting $\eta$ as
\begin{eqnarray}\label{EtaNLO eq}
\eta^{\mathrm{NLO}}{\simeq}
- \frac{1.55958}{\pi^2}\int_{\mu_0^2}^{\mu^2}
{\alpha^{2}_{s}(t)}\, d \ln t,
\end{eqnarray}
we find that the DIS entropy at the NLO approximation is now given by
\begin{eqnarray} \label{EntropyNLOs2 eq1}
{S}^{\mathrm{NLO}}(x,
\mu^2)&=&\ln{\bigg{[}}G^{\mathrm{NLO}}(x,\mu_{0}^2)
+\int_{x}^{1}G^{\mathrm{NLO}}(z,\mu_{0}^2)
\frac{d}{d({\ln}\frac{z}{x})}\mathrm{H}(\eta^{\mathrm{NLO}} {({\ln} \, \frac{z}{x}})^2)\frac{dz}{z}\bigg{]} .
\end{eqnarray}
The DIS entropy at the initial gluon distribution in the scale
$\mu_{0}^{2}$ is defined by the following form
\begin{eqnarray}\label{EntropyNLOQ0 eq}
S^{\mathrm{NLO}}(x,\mu_{0}^2){\simeq}{\ln}\bigg{[}xg^{\mathrm{NLO}}(x,\mu_{0}^2)\bigg{]}.
\end{eqnarray}

$\bullet$ $\mathbf{NNLO~ analysis}$:\\

At small $x$, we return to the end-point behavior of the three-loop
gluon-gluon splitting function in $s$-space as
\begin{eqnarray}
h^{(2)}(s)|_{s{\rightarrow}0}{\simeq}-\frac{E_{1}}{s^{2}}+\frac{E_{2}}{s},
\end{eqnarray}
where $E_{1}{\simeq}2675.85+157.269n_{f}$ and
$E_{2}{\simeq}14214.2+182.958n_{f}-2.79853n^{2}_{f}$ \cite{Vogt}.

Retaining the terms $1/s$  in the limit $s{\rightarrow}0$ and
rewriting $\xi$ as
\begin{eqnarray}\label{XiNNLO eq}
\xi^{\mathrm{NNLO}}{\simeq}
\int_{\mu_0^2}^{\mu^2}\bigg{[}\frac{3}{\pi}{\alpha_{s}(t)}-
\frac{61}{36\pi^2}{\alpha^{2}_{s}(t)}+\frac{232.832}{\pi^3}{\alpha^{3}_{s}(t)}\bigg{]}\,
d \ln t,
\end{eqnarray}
We find that the DIS entropy, depending on the initial conditions, is given at the NNLO approximation by
\begin{eqnarray} \label{EntropyNNLO eq}
{S}^{\mathrm{NNLO}}(x,
\mu^2)&=&\ln{\bigg{[}}G^{\mathrm{NNLO}}(x,\mu_{0}^2)+\int_{x}^{1}G^{\mathrm{NNLO}}(z,\mu_{0}^2)
\frac{\sqrt{\xi^{\mathrm{NNLO}}}}{\sqrt{{\ln}\frac{z}{x}}}\mathrm{BesselI}(1,2\sqrt{\xi^{\mathrm{NNLO}}}\sqrt{{\ln}\frac{z}{x}})\frac{dz}{z}\bigg{]}.
\end{eqnarray}

By retaining only the terms $1/s^2$  in the limit $s{\rightarrow}0$ and
rewriting $\eta$ as
\begin{eqnarray}\label{EtaNNLO eq}
\eta^{\mathrm{NNLO}}{\simeq}
\int_{\mu_0^2}^{\mu^2}\bigg{[}-
\frac{1.55958}{\pi^2}{\alpha^{2}_{s}(t)}+\frac{51.6395}{\pi^3}{\alpha^{3}_{s}(t)}\bigg{]}\,
d \ln t.
\end{eqnarray}
we find that the DIS entropy at the NLO approximation is now given by
\begin{eqnarray} \label{EntropyNNLOs2 eq1}
{S}^{\mathrm{NNLO}}(x,
\mu^2)&=&\ln{\bigg{[}}G^{\mathrm{NNLO}}(x,\mu_{0}^2)
+\int_{x}^{1}G^{\mathrm{NNLO}}(z,\mu_{0}^2)
\frac{d}{d({\ln}\frac{z}{x})}\mathrm{H}(\eta^{\mathrm{NLO}} {({\ln} \, \frac{z}{x}})^2)\frac{dz}{z}\bigg{]} .
\end{eqnarray}
The DIS entropy at the initial gluon distribution in the scale
$\mu_{0}^{2}$ is defined by the following form
\begin{eqnarray}\label{EntropyNNLOQ0 eq}
S^{\mathrm{NNLO}}(x,\mu_{0}^2){\simeq}{\ln}\bigg{[}xg^{\mathrm{NNLO}}(x,\mu_{0}^2)\bigg{]}.
\end{eqnarray}

$\bullet$ $\mathbf{Gluon~ distributions}$:\\

 The evolution of the DIS entropy in Eqs.~
(\ref{EntropyLO eq}),~(\ref{EntropyNLO eq}) and (\ref{EntropyNNLO
eq}) at the LO, NLO and NNLO approximations, respectively, depends
on the gluon distribution as described in Eqs.~ (\ref{EntropyLOQ0
eq}),~(\ref{EntropyNLOQ0 eq}) and (\ref{EntropyNNLOQ0 eq}) at the
initial scale $\mu_{0}^{2}$. Typically, parametrization groups
(such as  CT18 \cite{CT}, MSTW \cite{MSTW}, JR09 \cite{JR} and
NNPDF \cite{NNPDF1, NNPDF2} collaborations) use the following form
for the gluon distribution function at the initial scale
$\mu_{0}^{2}$:
\begin{eqnarray}\label{gluongeneral eq}
xg(x,\mu_{0}^{2})=A_{g}x^{\delta_{g}}(1-x)^{\eta_{g}}p_{g}(y(x)),
\end{eqnarray}
where $p_{g}$ is a widely used functional form in parton
distribution function sets, with the input $y(x)$ varying between
sets and being replaced by a neural network $NN_{g}(x)$
\cite{NNPDF1, NNPDF2}. Theoretical groups conduct global fits on
available experimental data to extract parton distribution
functions, which are crucial for studies involving colliding
hadrons. Currently, there are thousands of published data points
from various experiments that contribute to the extraction of more
precise parton distribution functions and strong coupling
constants at higher perturbative orders. For increased accuracy,
QCD fits can be conducted at LO, NLO, and even NNLO QCD
approximations. In this paper, we rely on the work of several
theoretical groups that provide global QCD fits to the gluon
distribution function  at
the initial scales using the following forms:\\
\begin{itemize}
\item[\tiny$\blacksquare$] The MSTW \cite{MSTW} at the LO approximation at the
input scale $\mu_{0}^{2}=1~\mathrm{GeV}^2$ (the NLO and NNLO
approximations according to Fig.1 are negative at low values of
$x$ ($x<0.01$).) reads
\begin{eqnarray}\label{MSTW eq}
xg(x,\mu_{0}^{2})=A_{g}x^{\delta_{g}}(1-x)^{\eta_{g}}[1+\epsilon_{g}
\sqrt{x}+\gamma_{g}x]+A_{g'}x^{\delta_{g'}}(1-x)^{\eta_{g'}}.
\end{eqnarray}
\item[\tiny$\blacksquare$] The CJ15 \cite{CJ} at the NLO approximation at the input
scale $\mu_{0}=m_{c}$ reads
\begin{eqnarray} \label{CJ15 eq}
xg(x,\mu_{0}^{2})=A_{g}x^{\delta_{g}}(1-x)^{\eta_{g}}[1+\epsilon_{g}
\sqrt{x}+\gamma_{g}x],
\end{eqnarray}
where the charm quark mass is defined as
$m_{c}=1.29^{+0.077}_{-0.053}~\mathrm{GeV}$ \cite{HZ,H1}.\\
\item[\tiny$\blacksquare$]  The CT18 \cite{CT} at the NNLO approximation at the
input scale $\mu_{0}=1.3~\mathrm{GeV}$ reads
\begin{eqnarray} \label{CT eq}
xg(x,\mu_{0}^{2})=A_{g}x^{\delta_{g}-1}(1-x)^{\eta_{g}}[\sinh(\epsilon_{g})
(1-\sqrt{x})^3+\sinh(\gamma_{g})3\sqrt{x}(1-\sqrt{x})^2+
(3+2\delta_{g})x(1-\sqrt{x})+x^{3/2},
\end{eqnarray}
where the input gluon distribution parameters are given in Table
I.\\
\begin{table}
\centering \caption{The parameter values are provided for three
parametrization groups.}\label{table:table1}
\begin{minipage}{\linewidth}
\renewcommand{\thefootnote}{\thempfootnote}
\centering
\begin{tabular}{|l|c|c|c|} \hline\noalign{\smallskip}
Parameters& MSTW LO  & CJ15 NLO & CT18 NNLO  \\
\hline\noalign{\smallskip}
$A_{g}$ &  0.0012216 & 45.542 & 2.690   \\
$\delta_{g}$ &  $-0.83657^{+0.15}_{-0.14}$ & $0.60307{\pm}0.031164$ & 0.531   \\
$\eta_{g}$ &  $2.3882^{+0.51}_{-0.50}$ & $6.4812{\pm}0.96748$ & 3.148   \\
$\epsilon_{g}$ &  $-38.997^{+36}_{-35}$ & $-3.3064{\pm}0.13418$ & 3.032   \\
$\gamma_{g}$ &  $1445.5^{+880}_{-750}$ & $3.1721{\pm}0.31376$ & -1.705   \\
$A_{g'}$ &  - & - & -   \\
$\delta_{g'}$ &  - & - & -   \\
$\eta_{g'}$ &  - & - & -   \\
\hline\noalign{\smallskip}
\end{tabular}
\end{minipage}
\end{table}
\end{itemize}

\section{Results}

To predict the entropy at higher-order approximations, it depends
on the gluon distribution at the initial scale $\mu_{0}^{2}$ and
the QCD cut-off $\Lambda$. The QCD cut-off parameter in the
modified minimal subtraction (MS) scheme \cite{MS} is determined
using the 4-loop expression for the running of $\alpha_{s}$ in
Ref.\cite{Bethke}. The world average value for
$\Lambda_{\overline{\mathrm{MS}}}$ is defined to be
\begin{eqnarray} \label{Lambda eq}
\Lambda^{n_{f}=4}_{\overline{\mathrm{MS}}}=(292{\pm}16)~\mathrm{MeV},
\end{eqnarray}
for $n_{f}=4$.\\
In Fig.1, the gluon distributions based on the parametrization
groups (i.e., MSTW \cite{MSTW}, CJ15 \cite{CJ}, and CT18
\cite{CT}) at the initial scales in a wide range of the Bjorken
values of $x$ are plotted. The gluon distributions based on the
MSTW \cite{MSTW} at the NLO and NNLO approximations are negative
at low values of $x$. In the following, we used the CJ15 \cite{CJ}
at the NLO
approximation and the CT18 \cite{CT} at the NNLO approximation.\\
A comparison of the initial gluon distributions at lower and
higher order corrections for the MSTW LO and CJ15 NLO sets of
parametrizations, along with their errors, is shown in Fig.2. The
dashed bands represent uncertainties on these gluon distributions.
It is observed that the error bounds at higher order corrections
are smaller than those at lower orders. In Fig.3, a comparison is
made between the different evaluations for the gluon entropy at
the LO and NLO approximations based on the MSTW and CJ15
respectively. It is plotted as a function of $x$ for virtualities
$\mu^2=2, 10$ and $100~\mathrm{GeV}^2$. Results are shown with and
without the rescaling variable as the behavior of evolution of the
gluon entropy with the rescaling variable is
in line with others (in the following the rescaling variable is used).\\
In Fig.4, we show the gluon entropy evaluated in this work based
on the parametrization groups.  The  extracted values are compared
with the H1 collaboration data \cite{H1A} as a function of the
average Bjorken $<x_{\mathrm{bj}}>$ measured in different average
squared momentum transfer $<\mu^2>$ ranges, obtained from
$\sqrt{s}=319~\mathrm{GeV}$ ep collisions. For this dataset, the
track pseudorapidities in the hadronic center-of-mass frame are
limited to the range $0<\eta^*<4$. The H1 data included total
errors from statistical and systematic uncertainties. There is a
good comparison  between the $S(x,\mu^2)$ predicted at the LO, NLO
and NNLO approximations and the entropy reconstructed from hadron
multiplicity at very small $x$. The resulting gluonic entropy
aligns well
with the hadronic entropy at  moderate values of $\mu^2$ in the interval of the statistical errors.\\
\begin{figure}
\includegraphics[width=0.55\textwidth]{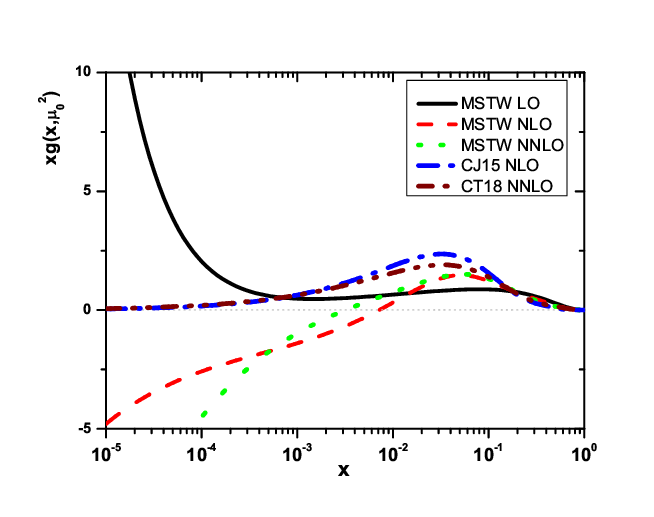}
\caption{The gluon distributions at the LO (solid-black), NLO
(dashed-red), and NNLO (dot-green) approximations from the MSTW
2008 PDFs \cite{MSTW} at the initial scale
$\mu_{0}=1~\mathrm{GeV}$, the  NLO (dashed-dot-blue) approximation
from the CJ15 PDFs \cite{CJ} at the initial scale $\mu_{0}=m_{c}$
and the  NNLO (dashed-dot-dot-brown) approximation from the CT18
PDFs \cite{CT} at the initial scale $\mu_{0}=1.3~\mathrm{GeV}$,
are plotted.}\label{Fig1}
\end{figure}
\begin{figure}
\includegraphics[width=0.55\textwidth]{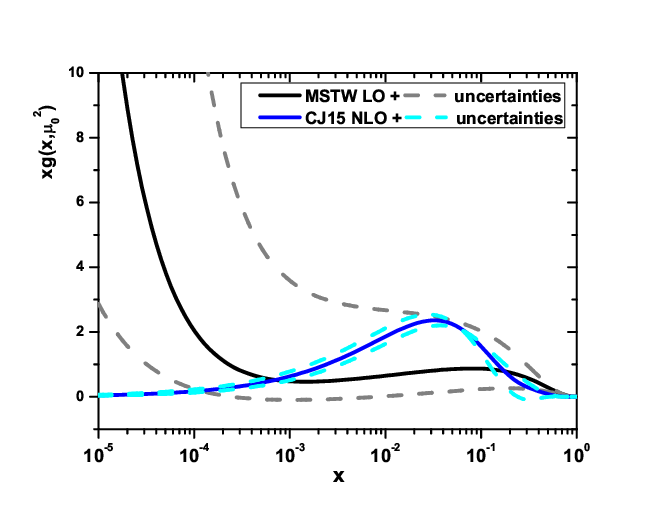}
\caption{The uncertainties in the comparison between the initial
gluon distributions at the LO  and  NLO  approximations from the
MSTW 2008 PDFs \cite{MSTW} and  the CJ15 PDFs \cite{CJ}  are
plotted.}\label{Fig2}
\end{figure}
\begin{figure}
\includegraphics[width=0.75\textwidth]{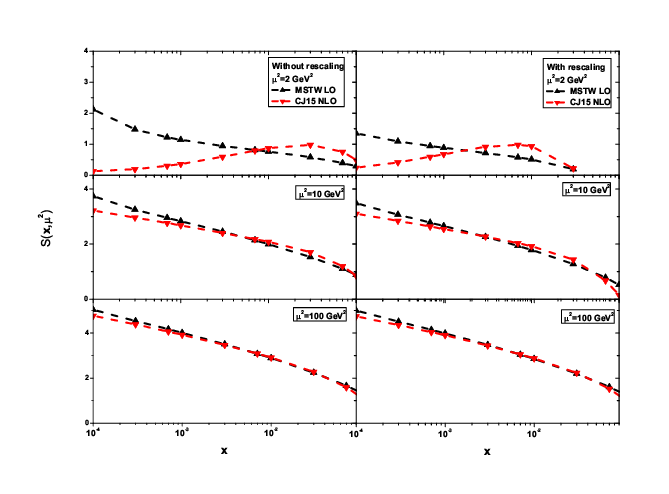}
\caption{ The evolution of gluonic entropy at $\mu^2=2, 10$ and
$100~\mathrm{GeV}^{2}$ without the rescaling (left diagrams) and
with the rescaling (right diagrams) based on the MSTW LO
\cite{MSTW} (solid black) and the CJ15 NLO \cite{CJ} (dashed
red).}\label{Fig3}
\end{figure}
\begin{figure}
\includegraphics[width=0.75\textwidth]{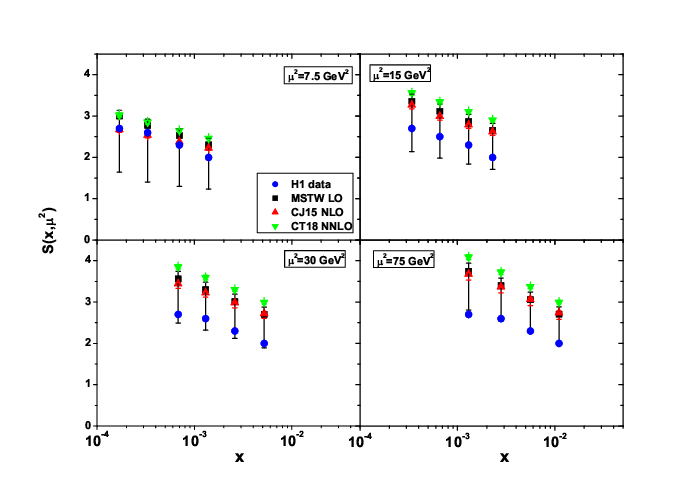}
\caption{ The evolution of gluon entropy is calculated using the
MSTW LO \cite{MSTW} (squared), the CJ15 NLO \cite{CJ} (up
triangle), and the CT18 NNLO \cite{CT} (down triangle) and
compared with H1 data \cite{H1A} as a function of $x$ at $\mu^2$
values of $7.5, 15, 30$, and $75~\mathrm{GeV}^2$. The H1
collaboration data \cite{H1A} is presented as a function of
$<x_{\mathrm{bj}}>$ measured in various averaged $\mu^2$ ranges at
$\sqrt{s}=319~\mathrm{GeV}$ ep collisions, accompanied by total
errors (For this dataset, the track pseudorapidities in the
hadronic center-of-mass frame are limited to the range
$0<\eta^*<4$ ).}\label{Fig4}
\end{figure}
The authors in Ref.\cite{Kutak3} have shown that the estimate of
charged versus total hadron multiplicity assumes that the total
number of produced hadrons is roughly $3/2$ times the number of
charged hadrons observed in experiments, as the partonic entropy
is defined by
\begin{eqnarray} \label{Entropyharged eq}
S_{\mathrm{Partonic}}{\rightarrow}S_{\mathrm{Charged}}=S_{\mathrm{Partonic}}+{\ln}(\frac{2}{3}).
\end{eqnarray}
In Table II, we display the charged results at the LO and NLO
approximations from Eq.~(\ref{Entropyharged eq})  and compare them
with the H1 hadron entropy derived from multiplicity distributions
measured in ep DIS at $<\mu^2>=30~\mathrm{GeV}^{2}$ as a function
of $<x>$. These results show that despite the very small
uncertainties in the H1 data, both corrections (i.e., LO and NLO)
provide very good fits to the data, relating the former to the
entropy of final-state hadrons.\\
\begin{table}
\centering \caption{The gluonic entropy, corrected for charged
hadrons (i.e., Eq.~(\ref{Entropyharged eq})), is only
${\ln}[xg]+{\ln}(\frac{2}{3})$ at $<\mu^2>=30~\mathrm{GeV}^{2}$.
This is compared in the LO and NLO approximations by the H1
collaboration data \cite{H1A}.}\label{table:table1}
\begin{minipage}{\linewidth}
\renewcommand{\thefootnote}{\thempfootnote}
\centering
\begin{tabular}{|l|c|c|c|} \hline\noalign{\smallskip}
$<x>$ & LO  & NLO & H1  \\
\hline\noalign{\smallskip}
$0.0052$ &  $2.297^{+0.17}_{-0.81}$ & $2.316^{+0.03}_{-0.12}$ & $2{{\pm}0.027}$   \\
$0.0026$ &  $2.604^{+0.18}_{-0.89}$ & $2.575^{+0.03}_{-0.12}$ & $2.3{{\pm}0.021}$   \\
$0.0013$ &  $2.895^{+0.18}_{-0.98}$ & $2.819^{+0.03}_{-0.11}$ & $2.35{{\pm}0.021}$   \\
$0.00068$ &  $3.154^{+0.18}_{-1.07}$ & $3.036^{+0.03}_{-0.11}$ & $2.4{{\pm}0.021}$   \\
\hline\noalign{\smallskip}
\end{tabular}
\end{minipage}
\end{table}
It is worth mentioning the pomeron intercept via scaling entropy
analysis that has been recently determined in the geometrical
scaling properties of the inclusive DIS cross section in
Ref.\cite{Machado}. This intercept is expressed in terms of the scaling
entropy obtained from event multiplicities $P(N)$ of final-state
hadrons, which is a more efficient way to detect scaling in
experimental data. In the Boltzmann-Gibbs (BG) form, the entropy is
given by
\begin{eqnarray}\label{BG eq}
S(x)=-\int
\mathcal{P}(x,k_{T})\ln[\mathcal{P}(x,k_{T})]d^{2}k_{T},
\end{eqnarray}
where $\mathcal{P}(x,k_{T})$ is the scattering amplitude in the
transverse momentum space\footnote{ In the dipole picture
\cite{Nikolaev}, the cross section in transverse momentum space
can be expressed as a convolution of the photon wave function
$|\psi_{\gamma}(k_{T},z)|^2$ with $\mathcal{P}(x,k_{T})$ (which
can be interpreted as a probability distribution containing all
information about the interaction process at the partonic level):
$$
\sigma_{\gamma^{*}p}(x,\mu^2)=\sigma_{0}\int{d^{2}k_{T}}dz|\psi_{\gamma}(k_{T},z)|^2\mathcal{P}(x,k_{T}).
$$
The form of $\mathcal{P}(x,k_{T})$ in the Tsallis entropy
\cite{Tsallis} is defined as follows:
$$
S_{q}=\int{d^{2}k_{T}}\frac{1-[\mathcal{P}(x,k_{T})]^q}{q-1},
$$
where the $q$-index represents the degree of nonextensivity of the
distribution. The Fourier transform of the scattering amplitude,
$\mathcal{P}(x,k_{T})$, is normalized to unity as $\int
{d^{2}k_{T}} \mathcal{P}(x,k_{T})=1$ with the constraint
$<k^{2}_{T}>_{q}=\beta^{-1}$ (where the Lagrange parameter $\beta
$ can be expressed in a scaling hypothesis
$<k^{2}_{T}>_{q}{\sim}\beta^{-1}(x_{x}/x)^\lambda$). Therefore the
probability distribution is defined in a scaling form
\cite{Machado}:
$$
\mathcal{P}(x,k_{T}){\sim}
\frac{1}{x^{-\lambda}}f(k^{2}_{T}/x^{-\lambda}).$$ Note that if
$\int{d^{2}k_{T}} \mathcal{P}(x,k_{T})=1$, then $\mathcal{P}$ has
dimensions of $1/\mathrm{mass}^2$. Therefore, $\mathcal{P}^q$ has
dimensions of $\mathrm{mass}^{-2q}$. Consequently $1-
\mathcal{P}^q$ is not a meaningful quantity. To make it
meaningful, it is necessary to introduce a mass scale $M$ and
consider $1- (\mathcal{P}/M)^q$.}. The entropy results, assuming
the scaling relation holds, are defined by the following form
\begin{eqnarray} \label{Slambda eq}
S=C+\lambda{\ln}(\frac{1}{x}).
\end{eqnarray}
The constant $C$, due to the power-like gluon distribution, can be
estimated into the hadron entropy as defined in
Ref.\cite{Machado}. In multiplicity data (the HERA ep data), the
entropy is defined as
\begin{eqnarray} \label{Mult eq}
S^{\mathrm{mult}}=-\sum_{N}P(N)\ln(P(N)),
\end{eqnarray}
where $P(N)$ is the probability of detecting $N$ charged hadrons.
The authors in Ref.\cite{Machado} obtained the averaged value of
$\lambda$ by scaling of the partonic entropy at each $\mu^2$ bin
(for $\mu^2=7.5,~ 15, ~30$ and $70~\mathrm{GeV}^2$) as
\begin{eqnarray} \label{Slambda value eq}
\lambda_{\mathrm{entropy}}=0.322~{\pm}~0.007.
\end{eqnarray}
In Fig.5, we show a calculation of the derivative
\begin{eqnarray} \label{SlambdaEq eq}
\bigg{(}\frac{{\partial}S(x,\mu^2)}{\partial{\ln}(1/x)}
\bigg{)}_{\mu^2}{\equiv}\lambda(x,\mu^2)
\end{eqnarray}
of the gluonic entropy $S(x,\mu^2)$ in the low $x$ domain of
deeply inelastic ep scattering \cite{H1L, Desg}. The behavior of
the determined values of $\lambda$ is presented due to the MSTW LO
\cite{MSTW}, the CJ15 NLO \cite{CJ}, and the CT18 NNLO \cite{CT}
in Fig.4 at $\mu^2=30~\mathrm{GeV}^2$. The curves in a wide range
of $x$ are compared by the scaling of the partonic entropy value
$\lambda_{\mathrm{entropy}}=0.322$ \cite{Machado}, by the
inclusive cross section method $\lambda_{\sigma}=0.329$
\cite{Machado} and the bCGC model \cite{Wat} which gives
$\lambda_{\mathrm{bCGC}}{\simeq}0.18$.\\
The values of $\lambda$ predicted in the literature are constant
as plotted in Fig.5, although $\lambda(x,\mu^2)$ depends on $x$.
Indeed, the evolution of entropy due to the running coupling order
is defined  by an effective intercept as \cite{Desg}
\begin{eqnarray} \label{Seffective eq}
\frac{{\partial}S(x,\mu^2)}{\partial{\ln}(1/x)}=\lambda(x,\mu^2)+\ln(\frac{1}{x})\frac{\partial{\lambda}(x,\mu^2)}{\partial{\ln}(1/x)}.
\end{eqnarray}
We observe that, in Fig.5, $\lambda$ depends on $x$. Therefore,
the effective intercept and $x$-slope do not coincide. We conclude
that one needs to be very careful when considering entropy and its
behavior in the small-$x$ region. In particular, at fixed $\mu^2$
and $x{\rightarrow}0$, we observe (in Fig.5) a decreasing
$x$-slope and hence $\lambda(x,\mu^2)$ at the higher-order
approximations. One can see in Fig.5 that the curve calculated in
the NLO approximation is close to the theoretical values of
$\lambda$ for $x{\leq}10^{-2}$.\\
The results at the NLO and NNLO approximations decrease as $x$
values decrease if we considered the $1/s^2$ terms\footnote{The
inverse Laplace transform of $\exp(A/s+B/s^2)$, where A and B are
constants, can be addressed in principle. However, there is not a
simple expression for the
result, just a double series:\\
\begin{eqnarray}
\mathcal{L} ^{-1}[e^{(a/s+b/s^2)} ; \upsilon] &=&
\delta(\upsilon)+ \sum_{n=1}^{\infty} \sum_{k=0}^{n} \frac{a^{n-k}
b^k}{k!(n-k)!(n+k-1)!}{ \upsilon} ^{n+k-1}.\nonumber
 \end{eqnarray}} in the
high-energy region of the coefficients $h^{(n)}_{gg}(s)$ as
determined in Eqs.~(\ref{EntropyNLOs2 eq1}) and
~(\ref{EntropyNNLOs2 eq1}). Indeed, such an evolution should lead
to a decreases of the growth of entropy at the higher-order
corrections. The estimate of $\lambda$ at the NLO approximation in
the interval $10^{-4}{\leq}x{\leq}10^{-2}$ at
$\mu^2=10~\mathrm{GeV}^2$ is
$<\lambda>_{\mathrm{NLO}}{\simeq}0.28$ which is in agreement with
the MM model \cite{Machado}.\\
\begin{figure}
\includegraphics[width=0.55\textwidth]{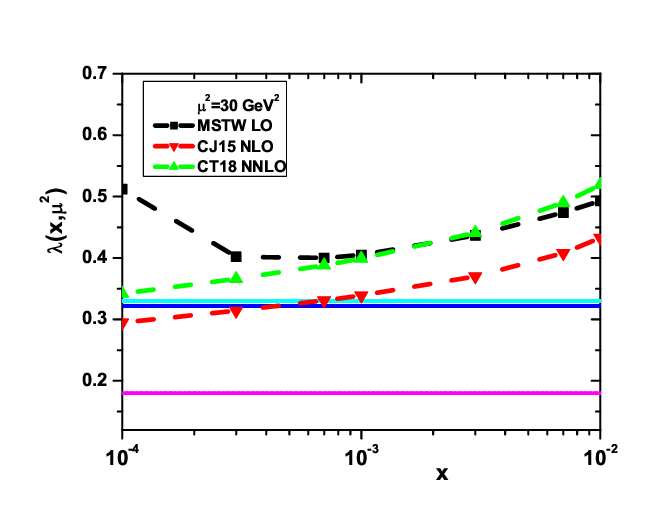}
\caption{ The values of $\lambda(x,\mu^2)$ are obtained as a
function of $x$ at $\mu^2=30~\mathrm{GeV}^2$ using the MSTW LO
\cite{MSTW} (dashed-black squared), the CJ15 NLO \cite{CJ}
(dashed-red down triangles), and the CT18 NNLO \cite{CT}
(dashed-green up triangles) and compared with
$\lambda_{\mathrm{entropy}}=0.322$ \cite{Machado} (solid-blue),
$\lambda_{\sigma}=0.329$ \cite{Machado} (solid-cyan) and
$\lambda_{\mathrm{bCGC}}{\simeq}0.18$ \cite{Wat}
(solid-magneta).}\label{Fig5}
\end{figure}
\begin{figure}
\includegraphics[width=0.55\textwidth]{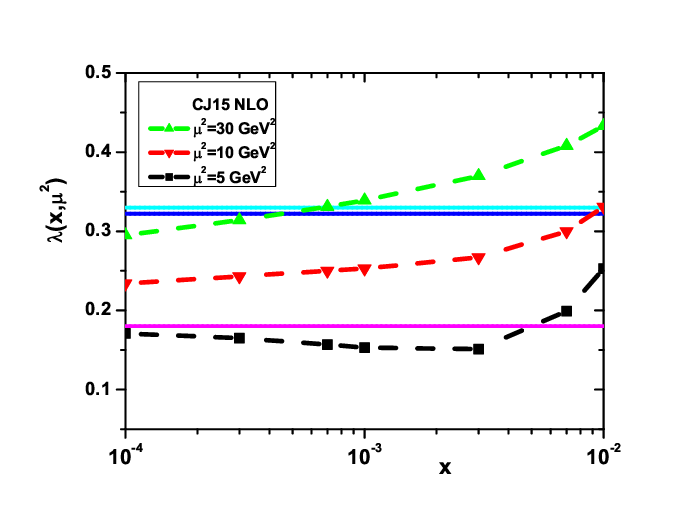}
\caption{ The values of $\lambda(x,\mu^2)$ are obtained as a
function of $x$ at $\mu^2=5, 10$ and $30~\mathrm{GeV}^2$ at the
NLO approximation using the CJ15 NLO \cite{CJ} (dashed-black
squared, dashed-red down triangles and dashed-green up triangles
respectively) and compared with $\lambda_{\mathrm{entropy}}=0.322$
\cite{Machado} (solid-blue), $\lambda_{\sigma}=0.329$
\cite{Machado} (solid-cyan) and
$\lambda_{\mathrm{bCGC}}{\simeq}0.18$ \cite{Wat} (solid-magneta).
}\label{Fig6}
\end{figure}
The average values of $\lambda$ in Fig.6 are approximately
independent of $\mu^2$ in the interval 0.18-0.33 at low values of
$x$. However, the $\lambda$ values depend on the $\mu^2$ values.
As shown $\lambda(x,\mu^2)$ increases as $\mu^2$ increases.
Therefore, entropy increases. The number of gluons and possibly
seaquarks that yield charged hadrons is effective in the $\lambda$
values obtained from the derivative of the DIS entropy, as
illustrated in Fig.6. Assuming the relationship between the
$x$-slope and ``the Pomeron effective intercept'' holds when the
entropy results in the following form \cite{Machado, Desg}:
\begin{eqnarray}\label{SPomeron eq}
S(x,\mu^2)=C(\mu^2)\bigg{(}\frac{1}{x}\bigg{)}^{\lambda(x,\mu^2)},
\end{eqnarray}
if one defines the Pomeron effective intercept as
$\alpha_{P}(x,\mu^2){\equiv}1+\lambda(x,\mu^2)$. At the same time
it seems possible, in principle, to derive some conclusions about
the effective intercept from the BFKL at the LO and NLO
approximations. The well known Balitsky-Fadin-Kuraev-Lipatov
(BFKL) Pomeron in the LO and NLO approximations has defined by the
following forms as \cite{Li1, Li2}
\begin{eqnarray}
\lambda_{\mathrm{BFKL}}^{\mathrm{LO}}=\alpha_{IP}-1=12{\ln{2}}(\alpha_{s}/\pi),
\end{eqnarray}
and
\begin{eqnarray}
\lambda_{\mathrm{BFKL}}^{\mathrm{NLO}}=\alpha_{IP}^{\overline{MS}}-1=12{\ln{2}}\frac{\alpha_{s}}{\pi}[1+r_{\overline{MS}}(0)\frac{\alpha_{s}}{\pi}],
\end{eqnarray}
where
$r_{\overline{MS}}(0){\simeq}-20.12-0.1020n_{f}+0.06692\beta_{0}$
and $\beta_{0}=\frac{1}{3}(33-2n_{f})$. We observe that the NLO
BFKL Pomeron intercept for $N_{C}=3$ and $n_{f}=4$ is calculated
to be $\lambda_{\mathrm{BFKL}}^{\mathrm{NLO}}{\simeq}-0.14$ for
$\alpha_{s}=0.2$. The results for the entropy with higher-order
corrections present an opportunity for utilizing
 NLO BFKL resummation in high-energy phenomenology.\\
 Recently, the authors in
Ref.\cite{Moriggi} have shown the values of $\lambda$ due to the
scaling entropy in charged hadron multiplicity distributions in
proton-proton collisions at the LHC. This dependence is on the
pseudorapidity interval $\eta$ in which the measurement is
performed and on the center-of-mass energy (COM) $\sqrt{s}$ of the
proton-proton collision. Assuming a minimal transverse momentum of
$p_T \sim 1~\mathrm{GeV}$ as the relevant scale for the partonic subprocess,
the corresponding Bjorken-$x$ value can be estimated using the following formula
\begin{eqnarray} \label{Rapidity eq}
x{\approx}\frac{e^{-\eta}}{\sqrt{s}}.
\end{eqnarray}
The $\lambda$ values at this limit can be expressed by the
following form
\begin{eqnarray} \label{LambdaRapidity eq}
\lambda(\sqrt{s},\eta){\equiv}\bigg{(}\frac{{\partial}S(\sqrt{s},\eta)}{\partial{\ln}(1/x)}
\bigg{)}_{x{\approx}\frac{e^{-\eta}}{\sqrt{s}}}.
\end{eqnarray}
The $\lambda$ values at the limit
$x{\approx}\frac{e^{-\eta}}{\sqrt{s}}$ in the NLO approximation
depend on the $\mu^2$ values, as shown in Fig.7. The average value
of pseudorapidity is assumed to be $<\eta>=1.5$, and the COM
energies are selected for the HERA and LHeC colliders as
$\sqrt{s}=319~\mathrm{GeV}$ and $1.3~\mathrm{TeV}$ respectively
\cite{Machado, LHeC}. The results are compared with the
$\lambda_{\mathrm{DIS}}=0.322{\pm}0.007$ \cite{Machado}, showing
that $\lambda$ rises approximately linearly (for
$\mu^2>10~\mathrm{GeV}^{2}$) with $\ln(\mu^2)$ as reported in
Ref.\cite{H1L} to the H1 structure
function data.\\
\begin{figure}
\includegraphics[width=0.55\textwidth]{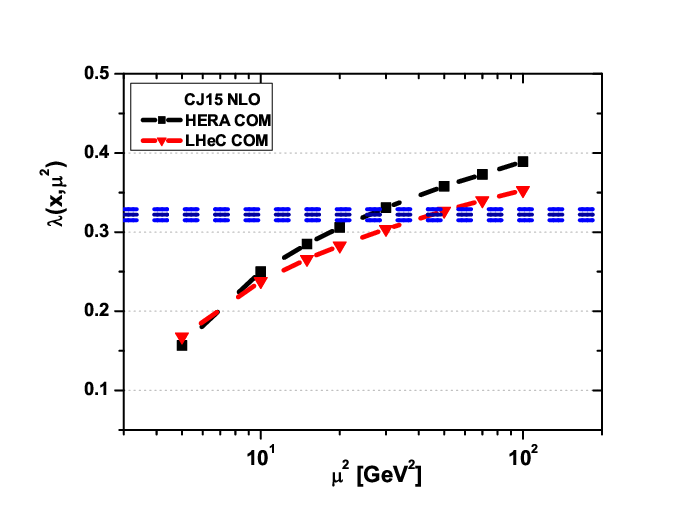}
\caption{ Extracted $\lambda$ values at pseudorapidity
$<\eta>=1.5$ and center-of-mass energies for the HERA and LHeC
colliders ($\sqrt{s}=319~\mathrm{GeV}$ and $1.3~\mathrm{TeV}$
respectively) compared to the
$\lambda_{\mathrm{DIS}}=0.322{\pm}0.007$ with the blue interval
corresponding to the H1 fit uncertainty. }\label{Fig7}
\end{figure}

\section{Conclusions}
We have presented a method based on the Laplace
transform to determine the evolution of gluonic entropy at
leading-order and higher-order approximations. This method relies
on the behavior of the gluon distribution function at initial
scales which depends on the running coupling.  The gluon
distributions at the initial scales are defined based on
parametrization groups of MSTW, CJ15, and CT18 at the LO, NLO,
and NNLO approximations, respectively. The results of the DIS
entropy with the rescaling variable are consistent with the H1
collaboration data reconstructed from hadron multiplicity at small $x$.\\
The gluonic entropy shows improvement with respect to the charged
hadron effects compared to the H1 hadron entropy, accompanied by
total errors. The Pomeron intercept via scaling entropy is
considered and compared with results obtained from event
multiplicities $P(N)$ of final-state hadrons. The behavior of
$\lambda(x,\mu^2)$ in the evolution of the DIS entropy with
respect to the running coupling is examined, showing dependence on
$x$. The values of $\lambda(x,\mu^2)$ decrease as the order of
evolution increases, which is consistent with the BFKL Pomeron in
the LO and
NLO approximations.\\
We conclude that the values of $\lambda(x,\mu^2)$ obtained from
scaling entropy fall within the range of results obtained from the
bCGC model and the inclusive cross-section of HERA data. For
$10^{-4}{\leq}x{\leq}10^{-2}$ and
$\mu^2{\geq}10~\mathrm{GeV}^{2}$, $\lambda$ is found to depend on
$x$ and to increase linearly with $\ln{\mu^2}$. We believe that
this investigation provides
insights into the dynamics of QCD at high energies.\\

%%%%%%%%%%%%%%%%%%%%%%%%%%%%%%%%%%%%%%%%%%%%%%%%%%%%%%%%%%%
\subsection{ACKNOWLEDGMENTS}

G.R.Boroun is grateful to Razi University for the financial
support provided for this project. Additionally, G.R.Boroun would
like to express thanks to Professor K.Kutak for his helpful
comments and invaluable support. Phuoc Ha would like to thank
Professor Loyal Durand for all his insightful comments and invaluable support.\\

%\subsection{Declaration of competing interest}
%The authors declare that they have no known competing financial
%interests or personal relationships that could have appeared to
%influence the work reported in this paper.\\

%\subsection{ Data availability}
% No data was used for the research described in the article.\\

%%%%%%%%%%%%%%%%%%%%%%%%%%%%%%%%%%%%%%%%%%%%%%%%%%%%%%%%%%%%%%%%%%%%%%%%%%

\end{document}